\documentstyle[aps,epsfig,floats]{revtex}
\newcommand{\bm}[1]{\mbox{\boldmath $#1$}}
\newcommand{\be}{\begin{equation}}
\newcommand{\ee}{\end{equation}}
\newcommand{\bea}{\begin{eqnarray}}
\newcommand{\eea}{\end{eqnarray}}
\newcommand{\st}{{\scriptscriptstyle T}}

\newcommand{\NP}[1]{{\it Nucl.\ Phys.}\ {\bf #1}}
\newcommand{\ZP}[1]{{\it Z.\ Phys.}\ {\bf #1}}
\newcommand{\PL}[1]{{\it Phys.\ Lett.}\ {\bf #1}}
\newcommand{\PR}[1]{{\it Phys.\ Rev.}\ {\bf #1}}
\newcommand{\PRL}[1]{{\it Phys.\ Rev.\ Lett.}\ {\bf #1}}

\newcommand{\EPJ}[1]{{\it Eur.\ Phys.\ J.}\ {\bf #1}}
\newcommand{\IJMP}[1]{{\it Int.\ J.\ Mod.\ Phys.}\ {\bf #1}}
\begin{document}
\tighten
\thispagestyle{empty}
\title{
\begin{flushright}
\begin{minipage}{4 cm}
\small
hep-ph/0001196\\ 
VUTH 00-04
\end{minipage}
\end{flushright}
\vspace{3mm}
Azimuthal Spin Asymmetries in Semi-Inclusive Production from Positron-Proton 
Scattering.
\protect} 
\vspace{3mm}
\author {M. Boglione and P.J. Mulders\\  
\vspace{3mm}
\mbox{}\\
{\it Division of Physics and Astronomy, Faculty of Science, Free University}\\
{\it De Boelelaan 1081, NL-1081 HV Amsterdam, the Netherlands}\\
{\sf boglione@nat.vu.nl, mulders@nat.vu.nl}
}

\maketitle

\vspace{1cm}

\begin{abstract}
The recent measurements of azimuthal single spin asymmetries by the HERMES 
collaboration at DESY may shed some 
light on presently unknown fragmentation and distribution functions.
We present a study of such functions and give some estimates of weighted 
integrals directly related to those measurements. 
\end{abstract}

\vspace{0.5cm}

PACS Numbers 13.85.Ni,13.87.Fh,13.88.+e

\section{Introduction}

The HERMES collaboration has recently presented interesting results on the 
measurement of some  single spin asymmetries, relative to inclusive 
pion production in the scattering of positrons off a longitudinally polarized 
hydrogen target~\cite{hermes99}. 
In particular, they are a $\sin(2\phi^l_h)$ and a $\sin(\phi^l_h)$ asymmetry, 
for which a theoretical analysis has been performed in 
Ref.~\cite{tm9596,bm98}.
Given as weighted cross-sections, $\langle W\rangle = \int W\,d\sigma$
with subscripts indicating polarization of beam and target, 
the relevant ones are

\bea
\left< \frac{Q_T^2}{4MM_h}\sin(2\phi ^l _h) \right>_{OL} =
- \frac{ 4 \pi \alpha ^2 s}{Q^4} \, \lambda (1-y)\,
\sum _{a,\overline a} e_a ^2 \ x_B \, h_{1L}^{\perp (1) a}
(x_B)  \, H_1 ^{\perp (1) a} (z_h) ,
\label{W1}
\eea

\bea
\lefteqn{
\left< \frac{Q_T}{M}\sin(\phi ^l _h) \right>_{OL} = 
\frac{ 4 \pi \alpha ^2 s}{Q^4} \,\lambda \,  (2-y)\sqrt{1-y}\,  
\frac{2M_h}{Q} } \nonumber \\ && \hspace*{1.0cm}
\times \sum _{a,\overline a} e_a ^2  
\Biggl\{ \, x_B \, h_{1L}^{\perp (1) a}(x_B) \, \frac{\tilde{H}^a(z_h)}{z_h} 
\, -  \, x_B \Bigl[x_B \, {h}_L^a(x_B) -\frac{m}{M} \, g_{1L}^a(x_B)\Bigr] 
\,H_1 ^{\perp (1) a} (z_h) \,
\Biggr\} ,
\label{W2}
\eea
where $\phi ^l _h$ is the azimuthal angle between the lepton scattering plane 
and the hadron production plane 
(see Ref.~\cite{bm98}), 
$M$ and $M_h$ are the masses of the target proton and of the produced 
hadron, respectively, whereas $Q_T$ is the transverse momentum of the 
produced hadron divided by $z_h$.
If we neglect the term proportional to the quark mass $m$ in 
Eq.~(\ref{W2}), 
we can see that four functions play a dominant role here: the distribution 
functions $h_{1L}^{\perp a}(x)$ and $h_L^a(x)$, 
and the fragmentation functions $H_1^{\perp a }(z)$ 
and  $\tilde H^{a}(z)$.
More precisely, the functions appearing in the weighted cross-sections are 
$h_{1L}^{\perp (1) a}(x)$ and  $H_1 ^{\perp (1)a}(z)$, where the superscript 
$(1)$ indicates that we are dealing with $k_T^2$-moments.
But let's examine our ingredients in some more detail.

The function
$h_{1L}^{\perp a}(x)$ is a leading (twist-two) chiral-odd distribution 
function, which describes the probability of finding a transversely polarized 
quark of flavour $a$ in a longitudinally polarized proton. 
The superscript $\perp$ 
signals a correlation between the proton longitudinal polarization, $\lambda$, 
and the intrinsic transverse momentum of the quark, $k_T$: the contribution to 
the correlator $\Phi_{ij}$ of the term proportional to $h_{1L}^{\perp}$ 
is zero any time we neglect intrinsic $k_T$. 

The function $h_L^a(x)$ is the twist-3 chiral-odd function relevant for 
a longitudinally polarized proton. 
It can be expressed in terms of leading functions 
plus interaction dependent terms as
\be
h_L(x) =
-\frac{2}{x} \, h_{1L}^{\perp (1)} (x)
+ \frac{m}{M\,x}\,g_1(x)
+ \tilde h_L (x) ,
\label{intermediate}
\ee
where $h_{1L}^{\perp (1)}$ is a $k_\st^2$-moment defined as
\be
h_{1L}^{\perp (1)}(x) = \int d^2k_\st
\ \frac{\vert \bm k_\st\vert^2}{2M^2}
\,h_{1L}^\perp (x,\bm k_\st).
\ee
By making use of a relation following from Lorentz covariance,
\be
h_L(x) = h_1(x) -\frac{d}{dx} h_{1L}^{\perp (1)} (x), 
\ee
one can solve Eq.~(3) for $h_{1L}^{\perp (1)}$ and obtain the well-known 
result~\cite{jj92}
\be
h_L(x) = 2x \, \int _x^1 dy\, \frac{h_1(y)}{y^2} 
\, + \,  \frac{m}{M} \Biggl(\frac{g_1(x)}{x} 
- 2x\int _x^1 dy \frac{g_1(y)}{y^3}\Biggr) 
\, + \, \Biggl(\tilde h _L (x) 
- 2x \int _x^1 dy \frac{\tilde h_L(y)}{y^2}\Biggr).
\label{h_Llong}
\ee
The last bracket in 
Eq.~(\ref{h_Llong}) contains the interaction dependent terms, involving  
$\tilde h_L (x)$, and will be indicated by $\overline h_L(x)$
\be
\overline h_L (x) = 
\tilde h _L (x) - 2x \int _x^1 dy \frac{\tilde h_L(y)}{y^2}\,.
\label{hL-bar}
\ee 
Neglecting the terms proportional to the quark mass $m$, one can simply
write $h_L(x)$ as
\be
h_L(x) = 2x \, \int _x^1 dy\, \frac{h_1(y)}{y^2} + \overline h_L (x)\,.
\label{h_Lshort}
\ee
Notice that $h_L(x)$, $\overline h_L (x)$ and $\tilde h_L(x)$, being higher 
twist, cannot be given an intuitive interpretation in terms of probability 
densities. 

As far as the fragmentation process is concerned, $H_1^{\perp}$ is a 
T-odd leading twist function which gives the probability of a spinless or 
unpolarized hadron (like the pion, for example) to be created from a 
transversely polarized 
scattered quark.
The role and the features of this function were extensively studied in 
Ref.~\cite{bm99} and in Ref.~\cite{abm99,bl99}, where
parameterizations based on a fit on $pp^{\uparrow} \to \pi X$ experimental 
data \cite{adams} was given.
It is worth to point out here that the contribution to the correlator 
$\Phi_{ij}$ of the term proportional to this function would be zero if the 
intrinsic transverse momentum of the fragmenting quark was neglected, as 
signaled by the superscript $\perp$.

Its first moment,  $H_1 ^{\perp (1) a} (z_h)$, which appears in the weighted 
integrals of Eqs.~(\ref{W1})~and~(\ref{W2}), is defined as
\be
H_1^{\perp (1)}(z) = \int d^2k_\st^\prime
\ \frac{\vert \bm k_\st\vert^2}{2M_h^2}
\,H_1^\perp (z,\bm k_\st^\prime).
\label{wH1Tperp}
\ee
%
%

The fragmentation function $\tilde H^a(z)$, appearing in the first term of 
Eq.~(\ref{W2}), is a subleading function which also can be split into
a leading function and an interaction dependent part,
\be
H^a(z) = 
-2z\,H_1^{\perp (1) a}(z) + \tilde H^a(z) .
\label{H-tilde}
\ee 
By making use of a relation following from Lorentz covariance,
\be
\frac{H^a(z)}{z}=z^2\,\frac{d}{dz}\,
\Bigl( \frac{H_1^{\perp (1) a}(z)}{z} \Bigr),
\ee
we can solve Eq. (10) to find: 
\be
\frac{\tilde H^a(z)}{z} = \frac{d}{dz}\,
\Bigl(zH_1^{\perp (1) a}(z) \Bigr),
\ee  
which straightforwardly connects $\tilde H^a(z)$ to $H_1^{\perp (1) a}(z)$.

Unfortunately, most of the distribution functions which 
appear in these expressions are not known a priori, since they have not been 
measured yet. Thus, no direct information can be extracted from the HERMES 
measurement.
Nevertheless, some light can be shed by considering extreme cases 
and exploit the consequences and the results they lead to. 
In what follows, we will examine in detail two possible opposite scenarios. 

It is worth mentioning that the HERMES collaboration also measured a third 
azimuthal single spin asymmetry, namely a $\sin\phi ^l _h$ asymmetry for polarized 
leptons. It can be expressed as a weighted integral as follows
\bea
\left< \frac{Q_T}{M_h}\;\sin(\phi ^l _h) \right>_{LO} = 
- \frac{ 4 \pi \alpha ^2 s}{Q^4} \, 2y\sqrt{1-y} \,
\sum _{a,\overline a} e_a ^2 \ \frac{M}{Q}\ x_B^2 \, 
\tilde{e}^a(x_B)  \, 
H_1 ^{\perp (1) a} (z_h) ,
\label{W3}
\eea 
in which, this time, it is the lepton beam to be polarized and not the 
hydrogen target.
This quantity involves, besides the same fragmentation function
as in the earlier-mentioned asymmetries,
the interaction dependent part of the higher twist distribution 
function $e^a(x)$, 
\be
e^a(x)
= 
\frac{m_a}{M}\frac{f_1^a(x)}{x}
+
\tilde{e}^a(x) .
\ee
The asymmetry in Eq. (\ref{W3}) is found to be small in HERMES experiment. 
Consistency among the various measurements seems to indicate that $\tilde 
e (x)$ is small.

\section{Results}

Our first approach is to assume that the contribution of the function 
$\tilde h _L (x)$, the interaction dependent term in $h_L$, and the
quark mass terms can be neglected.  This means  
\begin{eqnarray}
\overline h_L (x) &=& 0 , \\ 
h_L (x) &=& 2x \, \int _x^1 dy\, \frac{h_1(y)}{y^2} ,
\label{h_L}
\end{eqnarray}
as follows from Eqs.~(\ref{hL-bar}) and~(\ref{h_Lshort}). Furthermore, 
from Eq. (3) we find
\be
h_{1L}^{\perp (1)} (x) = -\frac{1}{2} x h_L(x) =
-x^2 \, \int _x^1 dy\, \frac{h_1(y)}{y^2} ,
\label{h1Lperp}
\ee
assuming suitable boundary conditions, $h_1(1)=0$.
Thus Eqs.~(\ref{h_L})~and~(\ref{h1Lperp}) allow us to express 
all the distribution functions we 
need in terms of one function only, the leading twist transverse spin 
distribution function $h_1(x)$. 
Very recently, this approximation has also been used in a calculation of the 
$\sin(\phi^l_h)$ and $\sin(2\phi^l_h)$ asymmetries using the effective 
chiral quark-soliton model \cite{e00}. 

\begin{figure}[t]
\begin{center}
\mbox{~\epsfig{file=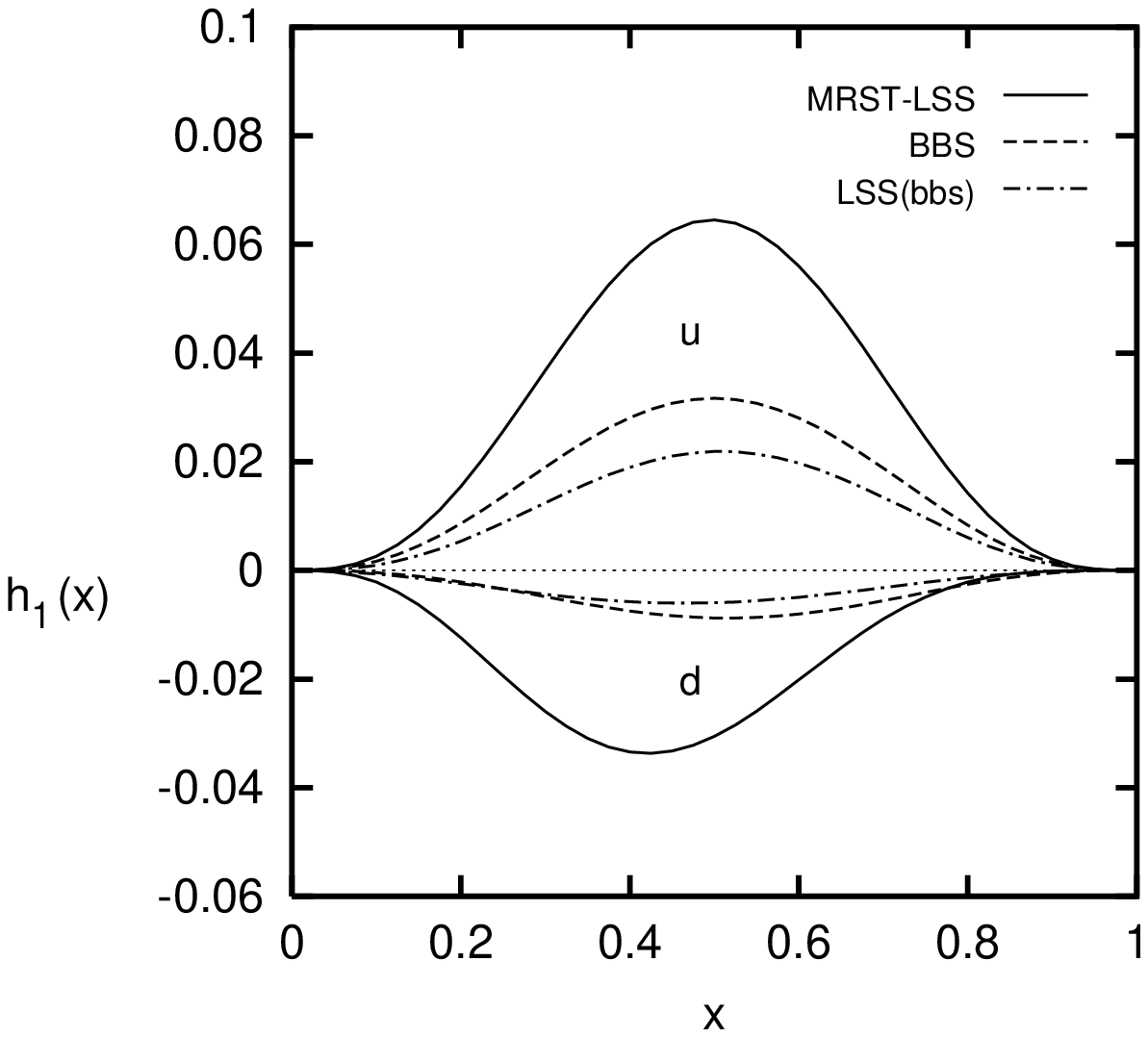,angle=-90,width=7.6cm}} 
\vspace{0.6cm}
\caption{\label{fig-h1}
The distribution functions $h_1^u(x)$ and $h_1^d(x)$ 
as obtained by using the  MRST-LSS, BBS and LSS$_{(BBS)}$ sets of distribution 
functions. The curves in the positive quadrant correspond to the $u$ flavour, 
whereas the curves in the negative quadrant correspond to the $d$ flavour.}
\end{center}
\end{figure}

As one possible input, we use the functions $h_1(x)$ and $H_1^{\perp}(z)$  
recently determined in Ref. \cite{bl99} by performing a new 
set of fits of the FNAL E704 $p^{\uparrow}p \to \pi X$ experimental data 
\cite{adams}. There, both the Soffer bound~\cite{soff} 
$|h_1(x)| \leq 1/2[f_1(x)+g_1(x)]$ and the positivity bound  
$H_1^{\perp}(z)\leq 2\,D_1(z)$ are respected, and it is showed how a 
completely satisfactory fit can only be obtained by using sets of 
distribution functions which respect the requirement $g_1/f_1 \to 1$ 
as $x \to 1$.
Strictly speaking, an unambiguous determination of these functions is not 
possible without  the aid of new and more accurate experimental data on a 
wider range, especially in the high $x$ region. Nevertheless, reasonable 
estimates can be given by using their parameterizations, which are the most 
involved and reliable presently available.     
Here, we will consider three of their choices of distribution functions: the 
old BBS parameterizations~\cite{bbs}, which 
respect the constraint $g_1/f_1 \to 1$ as $x \to 1$ and give the best fit 
in terms of $\chi^2$ in Ref. ~\cite{bl99} but does not involve any $Q^2$ 
evolution, the more recent LSS$_{(BBS)}$ set~\cite{lss-bbs}, parameterized 
in the same spirit but satisfying the correct $Q^2$ evolution and fitting 
the most recent world data and, for comparison, the 
LSS \cite{lss} and MRST \cite{mrst} sets of longitudinally polarized and 
unpolarized distribution functions, which include a 
``conventional'' $\Delta d(x)$, negative over the whole $x$ range.
Fig.~\ref{fig-h1} shows the function $h_1$ as obtained from the fit of 
Ref.~\cite{bl99} by using the three sets. 
Notice that the $h_1^u(x)$ and $h_1^d(x)$ obtained by 
using the BBS and LSS$_{(BBS)}$ distribution functions are roughly a 
factor 1/2 smaller than those obtained by using the LSS-MRST sets.

\begin{figure}[t]
\[\begin{array}{ll} \hspace*{-1.0cm}
\mbox{~\epsfig{file=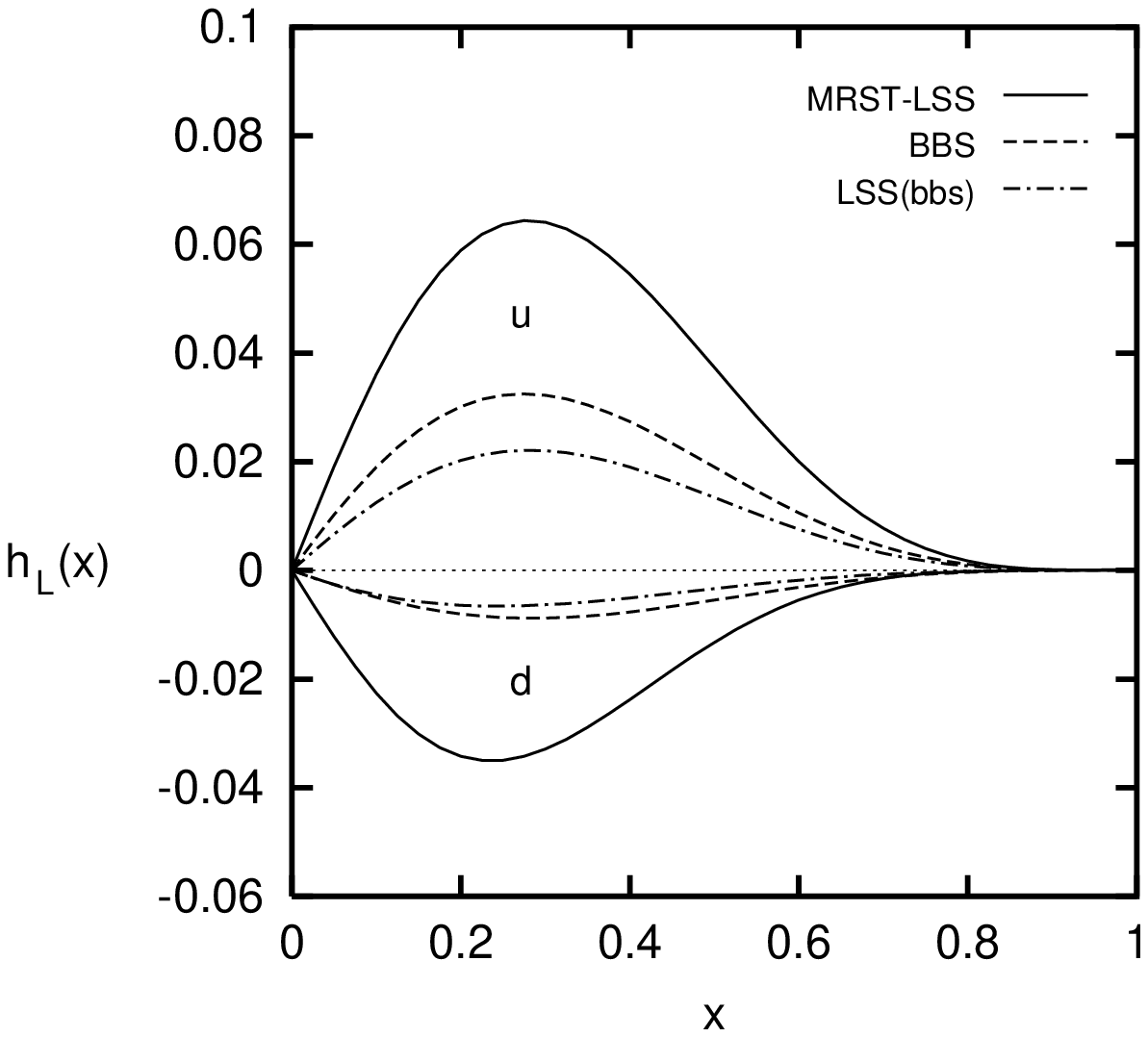,angle=-90,width=7.6cm}}
&
\mbox{~\epsfig{file=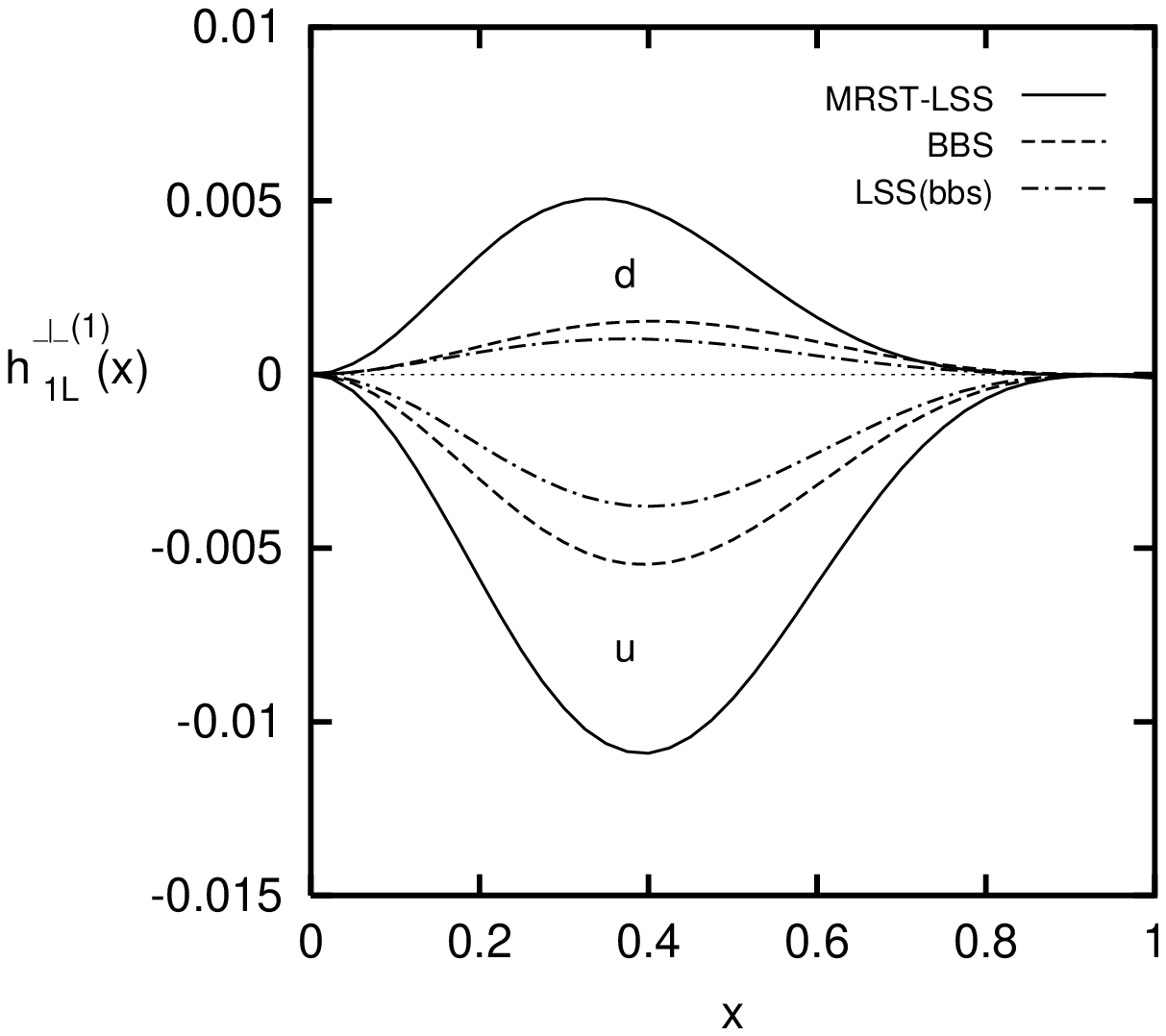,angle=-90,width=7.6cm}}
\vspace{0.2cm}
\end{array}\]
\caption{
\label{fig-hLperp}
The distribution functions $h_L^u(x)$, $h_L^d(x)$  and 
$h_{1L}^{\perp (1) u}(x)$, $h_{1L}^{\perp (1) d}(x)$, as obtained by using 
the MRST-LSS, BBS and LSS$_{(BBS)}$
sets of distribution functions respectively, 
under the approximation $\tilde h_L(x) =0$.}
\end{figure}

Substituting the explicit form of $h_1^a(x)$ in 
Eqs.~(\ref{h1Lperp}) and (\ref{h_L}), we can solve the integral and find the 
explicit parameterization of $h_{1L}^{\perp (1)a} (x)$ and $h_L ^a (x)$ 
(where $a$ means $u$ and $d$, since we are considering valence contribution 
only). 
The distribution functions $h_L(x)$ and $h_{1L}^{\perp (1)} (x)$ 
obtained assuming $\tilde h _L (x)=0$ in the two possible scenarios are 
presented in Fig.~\ref{fig-hLperp}. 
Notice that $h_{1L}^{\perp (1)} (x)$ satisfies the required bound 
$(h_{1L}^{\perp (1)} (x))^2 + (h_{1L}^{\perp} (x))^2 
\le \frac{p_T^2}{4M^2}f_1^2$, see Ref.~\cite{bbhm00} for details.  

The second assumption we consider is that $h_{1L}^{\perp (1)} (x)$ 
is small enough to be neglected (and again quark mass terms are
neglected too). It is interesting to point out that this approximation 
seems at first sight the most appropriate, since the HERMES collaboration 
finds the $\sin(2\phi _h ^l)$ single spin asymmetry of Eq.~(\ref{W1}) to be 
much smaller than the $\sin(\phi_h^l)$ asymmetry~\cite{hermes99}. 
A preliminary HERMES analysis \cite{oga} is actually going to use this 
approximation.  
We will comment on this choice later.

In this approximation, by using Eqs.~(\ref{intermediate}) and 
(\ref{h_Lshort}), we obtain 
\begin{eqnarray}
h_L(x) &=& h_1(x) \,, \\
\tilde h_L(x) &=& h_1(x) \,,\\
\overline h_L(x) &=& h_1(x) - 2x\int _x^1 dy \frac{h_1(y)}{y^2}\,.
\end{eqnarray}
Again, we can solve the integral and 
find an explicit parameterization of $h_L(x)$.
  
\begin{figure}[t]
\begin{center}
\mbox{~\epsfig{file=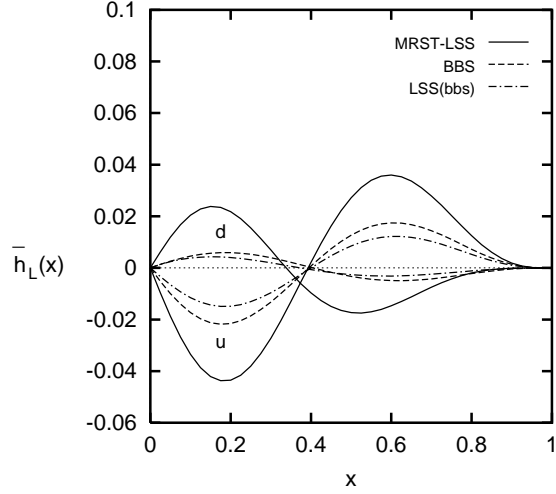,angle=-90,width=7.6cm}}
\vspace{0.6cm}
\caption{\label{fig-hLbar}
The distribution functions $\overline h_L^u(x)$ and $\overline h_L^d(x)$, 
as obtained  under the approximation $h_{1L}^{\perp (1)}(x)=0$, by using 
the  MRST-LSS, BBS and LSS$_{(BBS)}$ sets of distribution functions.}
\end{center}
\end{figure}

Plots of the distribution function $\overline h_L^a(x)$,  
obtained assuming $h_{1L}^{\perp (1)} (x)=0$ by the BBS, LSS$_{(BBS)}$ or the 
LSS-MRST sets of distribution functions, are presented in 
Fig.~\ref{fig-hLbar}. 
Notice that in both cases we find
\be
\int _0 ^1  \; dx \, h_L^a(x) = 0 \,.
\ee

To be able to calculate the weighted integrals in 
Eqs.~(\ref{W1},\ref{W2},\ref{W3}), 
we need an estimate of the fragmentation functions involved. 
$H_1^{\perp (1)}(z)$ was extensively studied and discussed in 
Refs.~\cite{bm99,abm99}, and in the recent Ref.~\cite{bl99} 
a suitable parameterization was given  which respects the positivity 
constraint and is consistent with the transversity distribution function 
$h_1(x)$ used above [see Ref.~\cite{bl99} for details and discussion]. 
We then have
\be
H_1^{\perp (1)}(z) = \frac{1}{z^{0.73}}\, \Big[\,1.21 \,(1-z)^{1.40} +
1.35\,(1-z)^{4.97}\,\Big] \;,
\label{H1perp}
\ee 
where we used the unpolarized pion fragmentation functions as given by 
Ref.~\cite{bkk1}, using isospin simmetry to separate the $\pi^+$ and $\pi^-$ 
contributions. The function
$\tilde H (z)$ can be expressed as a function of $H_1^{\perp (1)}(z)$ 
via Eq.~(\ref{H-tilde})
\bea
\tilde H(z) &=&
[0.33  \, (1 - z)^{1.40} + 0.37 \, (1 - z)^{4.97}] \, z^{0.27} + \\ &&
[-1.70 \, (1 - z)^{0.40} - 6.73 \, (1 - z)^{3.97}] \, z^{1.27} \;.
\eea 

One might be tempted to examine the two possible extreme situations, 
$H_1^{\perp (1)}=0$ or $\tilde H(z)=0$, in analogy to what was done for the 
distribution functions. But this would not lead to relevant results. 
In fact, if $H_1^{\perp (1)}=0$ then also $\tilde H(z)=0$ and all the 
weighted integrals would be zero. 
On the other hand, if $\tilde H(z)=0$, 
then Eq.~(\ref{H-tilde}) give the constraint $z\,H_1^{\perp (1)}=$ const, 
which is only consistent with the requirement of $H_1^{\perp (1)}$ being zero 
itself.  

\begin{figure}[p]
\begin{center}
\mbox{~\epsfig{file=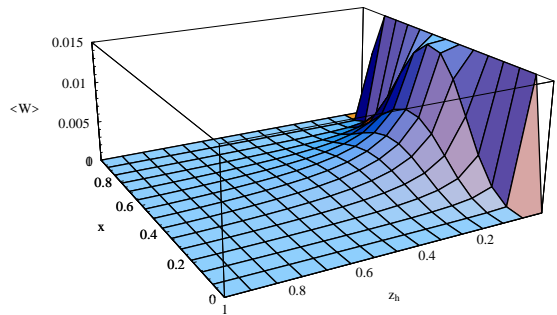,angle=0,width=7.5cm}}
\vspace{0.6cm}
\caption{\label{fig-WW1}
A three-dimensional view of 
$-\sum _{a,\bar a} e^2_a x_B h_{1L}^{\perp(1)a} (x_B) 
H_1  ^{\perp (1) a} (z_h)$, relevant for the $\sin (2\phi^l _h)$ asymmetry in 
$\pi^+$ production, under the approximation 
$\tilde h_L(x)= \overline h_L =0$, as obtained by using the BBS set of 
distribution functions. 
}
\end{center}
\end{figure}

\begin{figure}[p]
\[\begin{array}{ll} \hspace*{-1.0cm}
\mbox{~\epsfig{file=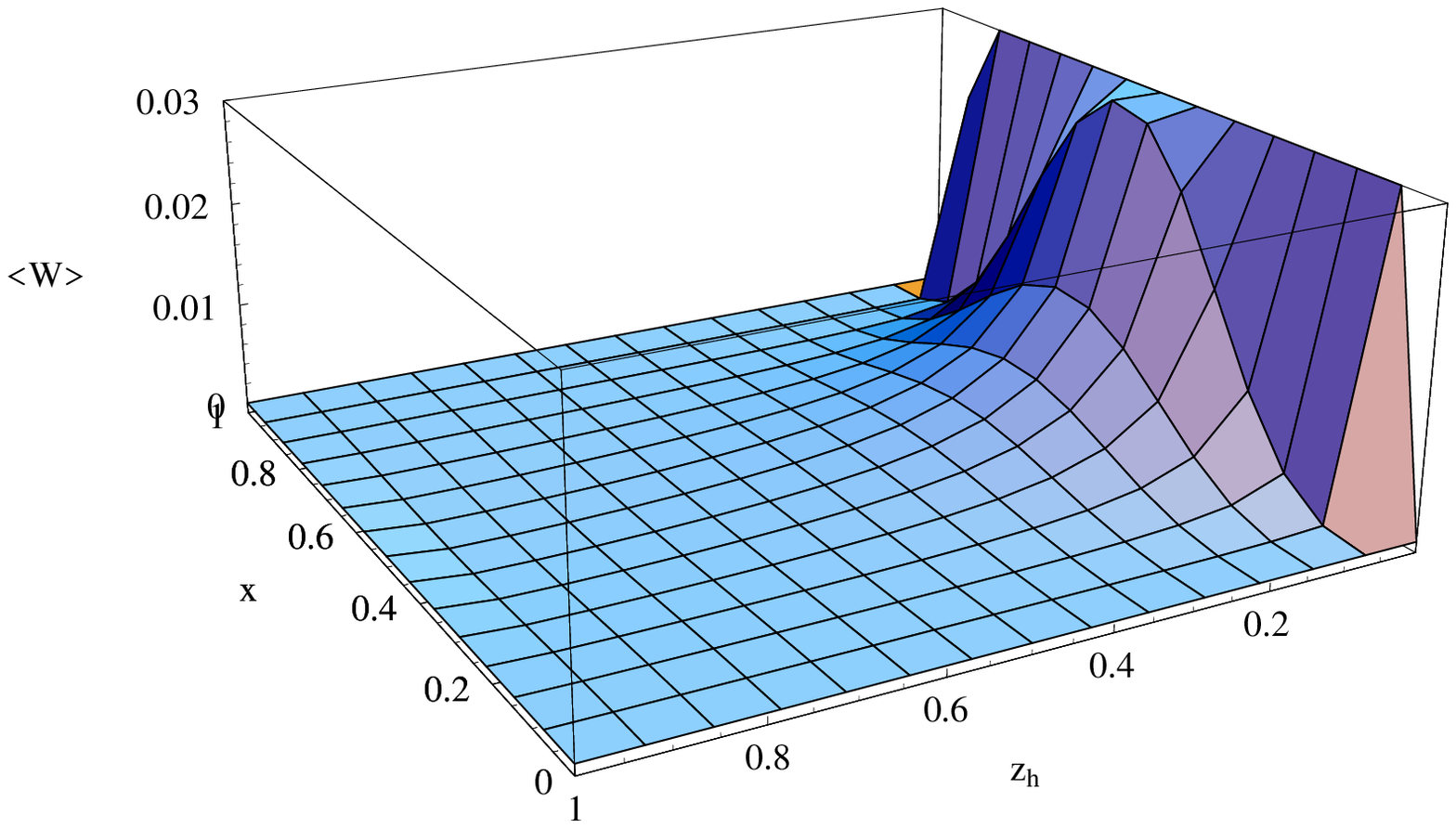,angle=0,width=7.5cm}}
&
\mbox{~\epsfig{file=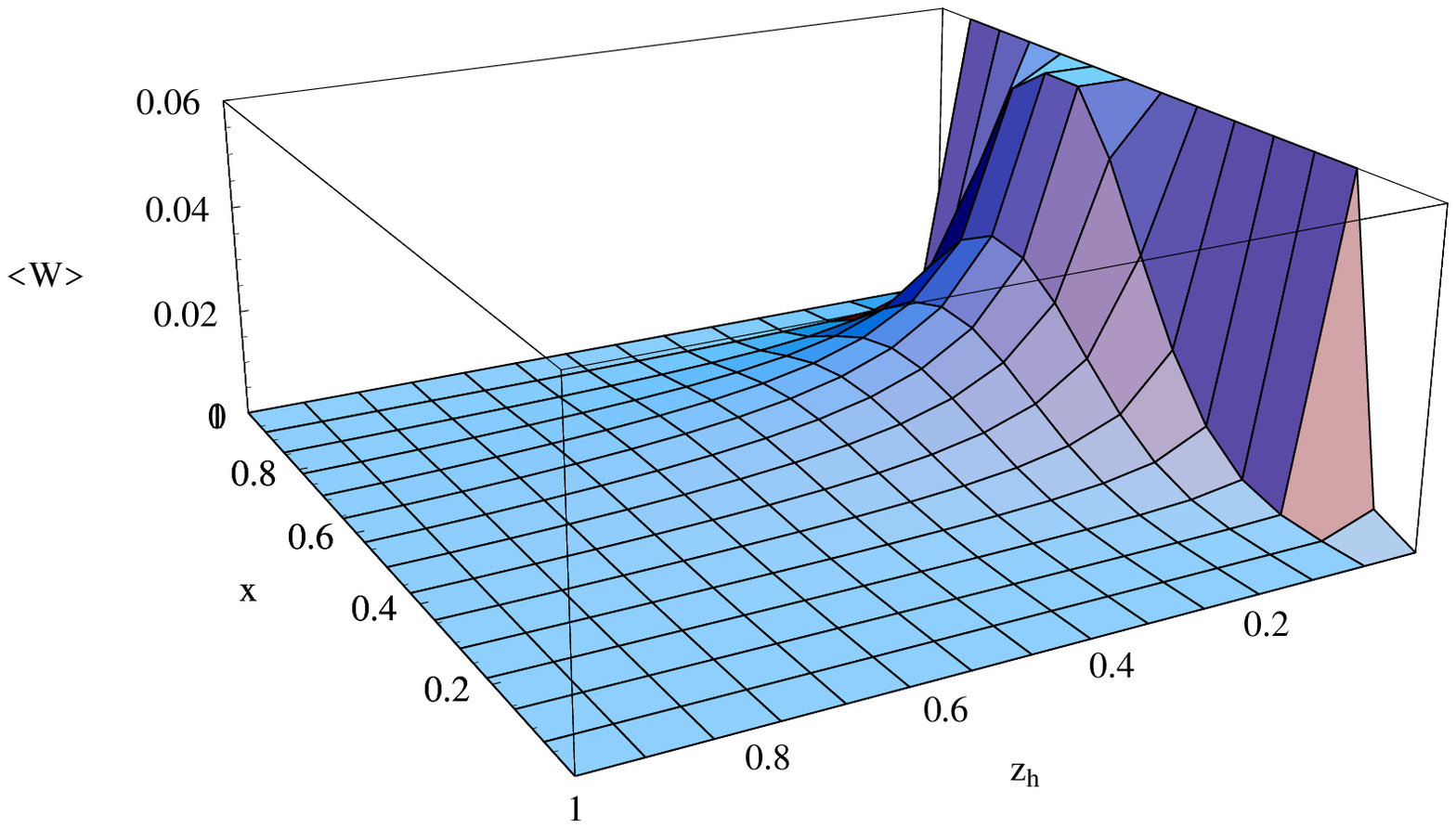,angle=0,width=7.5cm}}
\end{array}\]
\vspace{0.6cm}
\caption{
\label{fig-WW2-3}
A three-dimensional view of  
$-\sum _{a,\bar a} e^2_a \Bigl[x_B h_{1L}^{\perp(1)a} (x_B) 
\tilde H ^a (z_h)/z 
- x_B^2  h_L^a (x_B) H_1^{\perp (1) a}(z_h)\Bigr]$, relevant for the 
$\sin (\phi^l _h)$ asymmetry in $\pi^+$ 
production, as obtained by 
using the BBS set of distribution functions,
under the approximation $\tilde h_L = \overline h_L = 0$ (on the left) and 
under the approximation $h_{1L}^{\perp (1)}(x)=0$ (on the right).}
\end{figure}

In Figs.~\ref{fig-WW1} and \ref{fig-WW2-3} we present plots of the 
azimuthal spin asymmetries as a function of $x$ 
and $z$, obtained by using the BBS set of distribution 
functions. Choosing the LSS$_{(BBS)}$ set would give very similar results,
whereas for the MRST-LSS set the asymmetries retain the same shape and 
features but are larger of roughly a factor two. 
Fig.~\ref{fig-WW1} shows the weighted integral of 
Eq.(\ref{W1}) in the only scenario in which it is non-zero, i.e. for 
$\tilde h_L(x)=0$. The plots correspond to the BBS  
choice of distribution functions.
The weighted integral of Eq.~(\ref{W2}), corresponding to the two possible 
extreme situations we discussed in the previous session, is shown in 
Fig.~\ref{fig-WW2-3}.
Notice that under the approximation $\tilde h_L = \overline h_L = 0$, 
Fig.~\ref{fig-WW2-3}, both the terms proportional to  
$h_{1L}^{\perp (1)}$ and $h_L$ contribute to the weighted 
integral, 
whereas under the assumption $h_{1L}^{\perp (1)}(x)=0$ the weighted integral 
is proportional to  the term $h_L$ only.
It is interesting to notice that the $\sin(2\phi^l_h)$ asymmetry is suppressed 
compared to the $\sin(\phi^l_h)$ asymmetry  even in the approximation 
$\tilde h_L = 0$, which leads to a maximal $h_{1L} ^{\perp (1)}$. 
This tells us that the experimental measurement of HERMES yielding a small 
$\sin(2\phi^l_h)$ spin-asymmetry, consistent with zero, allows no 
conclusions on $h_{1L} ^{\perp (1)}$. This result is confirmed by the 
calculation in Ref.~\cite{e00}.
Note also that all the weighted integrals have roughly the same overall 
shape. They are sizeable in the small $z$ region for central values of $x$.  
Of course one needs to be aware that, depending on $Q^2$, at small $z$-values 
threshold effects in the production of hadrons and contributions from target 
fragmentation become important. 

\section{Conclusions}

Distribution and fragmentation functions are a fundamental issue. 
They tell us about the internal structure of the nucleons and of the role 
their elementary constituents play in accounting for their total spin.
It is then crucial to study those processes in which these functions can be 
exploited. After many years of efforts, both on the experimental and 
theoretical point of view, experimental information on polarized distribution 
and fragmentation functions is now starting 
to come from different sources (HERMES, SMC, SLAC, COMPASS and JLAB). Thus, 
some light can be shed, even though we are still far from a completely clear 
picture.
In this paper, we have studied two possible scenarios corresponding to two 
extreme approximations. Further experimental results could possibly give 
us enough handles to distinguish between the two extreme cases, and present 
more conclusive results and parameterization for the functions we would like 
to uncover.
This would be another step helping to draw a neater picture of the very 
intriguing ``soft'' physics which governs the hadronic world.

\section*{Acknowledgments}

\noindent
This work is part of the research program of the foundation for the 
Fundamental Research of Matter (FOM) and the TMR program ERB FMRX-CT96-0008.

\newpage

\listoffigures

\newpage

\newpage

\newpage

\newpage

\newpage

\end{document}